\documentclass[
    ,final            
  ]
  {aipproc}
\usepackage{amssymb}
\usepackage{amsmath}

\def\PD#1#2{\frac{\partial #1}{\partial #2}}

\def\BS#1{{\bf #1}}
\layoutstyle{8x11double}
\begin{document}
\title{Event-Driven Simulation of the Dynamics of Hard Ellipsoids}
\author{Cristiano~{De Michele}}{%
 address={Dipartimento di Fisica and INFM-CRS Soft,
         Universit\`a di Roma \emph{La Sapienza}, P.le A. Moro 2, 00185 Roma, Italy}
}
\author{Rolf~Schilling}{%
  address={Johannes-Gutenberg-Universitat Mainz, D-55099 Mainz,
Germany}
}
\author{Francesco~Sciortino}{%
 address={Dipartimento di Fisica and INFM-CRS Soft,
 Universit\`a di Roma \emph{La Sapienza}, P.le A. Moro 2, 00185 Roma, Italy}%
}%
\begin{abstract}
We introduce a novel algorithm to perform event-driven simulations of hard rigid bodies of arbitrary shape, 
that relies on the evaluation of the geometric distance.
In the case of a monodisperse system of uniaxial hard ellipsoids,
we perform molecular dynamics simulations varying the aspect-ratio $X_0$ and the packing fraction $\phi$.
We evaluate the translational $D_{trans}$ and the rotational $D_{rot}$ diffusion coefficient  and the
associated isodiffusivity lines in the $\phi-X_0$ plane.  We observe a decoupling of the translational and 
rotational dynamics which generates an almost perpendicular crossing of  the $D_{trans}$ and $D_{rot}$ isodiffusivity lines.   While the self intermediate scattering function exhibits stretched relaxation, i.e. glassy dynamics, only for large $\phi$ and $X_0 \approx 1$,  the second order orientational correlator $C_2(t)$ shows
stretching only for large and small $X_0$ values. We discuss these findings in the context of a possible pre-nematic order driven glass transition. 
%
%
%
\end{abstract}%
\maketitle
\section{Introduction}
Particles interacting with only excluded volume interaction may exhibit a rich
phase diagram, despite the absence of any attraction.  
Simple non-spherical hard-core particles can form  either crystalline or liquid crystalline ordered phases \cite{allenReview},
as first shown analytically  by Onsager \cite{Onsager} for rod-like particles. 
Although detailed phase diagrams of several hard-body (HB) shapes can be found in literature 
\cite{Parsons,Lee2,allenHEPhaseDiag,MargoEvans}
 less detailed information are available about dynamics properties of hard-core bodies and their kinetically arrested states.

The slowing down of the dynamics of the hard-sphere system on increasing the packing fraction $\phi$ 
is well described by mode coupling theory (MCT) \cite{GoetzeMCT}, but on going from spheres to non-spherical particles, 
non-trivial phenomena arise, due to the interplay between translational and rotational degrees of freedom.  The slowing down of the dynamics can indeed
appear either in both  translational and  rotational properties or in just one of  the two. 
Hard ellipsoids (HE) of revolution \cite{allenReview,singh01} are one 
of the most prominent systems composed by  hard body anisotropic particles.

The equilibrium phase diagram, evaluated numerically two decades ago \cite{FrenkelPhaseDiagMolPhys}, shows an isotropic fluid phase (I) and several ordered phases
(plastic solid, solid, nematic N). The coexistence lines show a swallow-like dependence with a minimum at the spherical limit $X_0=1$ and  a maximum at $X_0\approx 0.5$ and $X_0 \approx 2$ (cf. Figure \ref{Fig:isod}).
Application to HE \cite{LetzSchilLatz} of the molecular MCT (MMCT) \cite{A1ng,A2ng} predicts also  a swallow-like glass transition line. 
In addition, the theory suggests that for $X_0 \lessapprox 0.5$ and $X_0 \gtrapprox 2$, the glass transition is driven by a precursor of nematic order, resulting in an orientational glass where the translational density fluctuations are quasi-ergodic, except for very small wave vectors $q$.
Within MCT, dynamic slowing down associated  to a glass transition is driven by the amplitude of the static correlations.
Since the approach of the nematic transition line is accompanied by an increase of the nematic order correlation function at $q=0$, the non-linear feedback mechanism of MCT results in a glass transition, already before macroscopic nematic order occurs \cite{LetzSchilLatz}.  In the arrested state, rotational motions become hindered.

We perform event-driven (ED)  molecular dynamics simulations, using a new algorithm \cite{DeMicheleHEJCP,DeMichelePRL2007},
that differently from previous ones~\cite{AllenFrenkelDyn,DonevTorqStill1}, relies on the evaluation of the distance between objects of arbitrary shape and that will be illustrated shortly.

The outline of the manuscript is as follows. In the next section we illustrate shortly the new algorithm, that we proposed for simulating objects of arbitrary shape.
Then in Sec. \ref{Sec:methods} we illustrate the model used and we give all the details of the simulations we performed.
In Sec. \ref{Sec:results} we show the results concerning the dynamics of the system investigated.
The final section contains our conclusions.
\section{Algorithm details}
\label{Sec:algodetails}
\subsection{An event-driven algorithm for hard rigid bodies}
\label{sec:rbED}
In an ED simulation the system is propagated until the next event occurs, where an event can 
be a collision between particles, a cell crossing (if linked lists are used), etc. All these events must be ordered with respect to time in a calendar and insertion, deletion and retrieving of events must be performed as efficiently as possible.

One elegant approach has been introduced twenty years ago by Rapaport \cite{RapaBook}, who proposed to arrange all events into an ordered binary tree (that is the calendar of events), so that insertion, deletion and retrieving of events can be done with an efficiency $O(\log N)$, $O(1)$ and $O(\log N)$
respectively, where $N$ is the number of events in the calendar.  We adopt this solution to handle the events in our simulation; all the details of this method can be found in \cite{RapaBook}.
Our ED algorithm can be schematized, as follows:
\begin{enumerate}
\item Initialize the events calendar (predict collisions, cells crossings, etc.).
\item Retrieve next event $\cal E$ and set the simulation time to the time of this event. 
\item If the final time has been reached, terminate.
\item If $\cal E$ is a collision between particles $A$ and $B$ then:
\begin{enumerate}
\item change angular and center-of-mass velocities of $A$ and $B$ (see \cite{DonevTorqStill2}).
\item remove from the calendar all the events (collisions, cell-crossings) in which $A$ and $B$ are involved.
\item predict and schedule all possible collisions for $A$ and $B$.
\item predict and schedule the two cell crossings events for $A$ and $B$.
\end{enumerate}
\item If $\cal E$ is a cell crossing of a certain rigid body $A$:
\begin{enumerate}
\item update linked lists accordingly.
\item remove from calendar all events (collisions, cell-crossings) in which $A$ is involved.
\item predict and schedule all possible collisions for $A$ using the updated linked lists.
\end{enumerate}
\item go to step 2.
\end{enumerate}
All the details about linked lists can be found again in Ref. \cite{RapaBook}, where an ED algorithm for hard spheres is 
illustrated.
We note also that in the case of HE, according to \cite{DonevTorqStill2}, if the elongation of HE is big
(i.e. one axes is much greater or much smaller than the others), the linked list method becomes inefficient at
moderate and high densities.
In fact, given one ellipsoid $A$, using the linked lists, collision times of $A$ with all ellipsoids in the same cell of $A$
and with all the $26$ adjacent cells have to be predicted. In the case of rotationally symmetric ellipsoids,
the number of time-of-collision predictions grows as $X_0^2$, if $X_0 > 1$
and as $X_0$, if $X_0 < 1$ \cite{DonevTorqStill2}.
To overcome this problem, we developed also a new nearest neighbours list (NNL) method. 
At the begin of the simulation we build an oriented bounding parallelepiped (OBP) around each ellipsoid and we only predict collisions  between ellipsoids having overlapping 
OBP. In other words given an ellipsoid $A$, all the ellipsoids having overlapping OBP are the NNL of $A$.
In addition the time-of-collisioni $t_i$ of each ellipsoid with its corresponding OBP is evaluated
and the NNL \footnote{This is a particular case of the collision between rigid bodies.}
of each HE is rebuilt at the time $t=\min_i\{t_i\}$. 
The most time-consuming step in the case of hard rigid bodies is the prediction of collisions between HE, 
hence in the following a short description of the algorithm, that we developed, is given. 
\subsection{Distance between two rigid bodies}
Our algorithm for predicting the collision time of two rigid bodies relies on the evaluation of the distance
between them.
The surfaces of two rigid bodies, $A$ and $B$, are implicitly defined by the following equations:
\begin{subequations} \label{Eq:shapes}
\begin{equation}\label{Eq:shapef}
f({\bf x}) = 0
\end{equation}
\begin{equation}\label{Eq:shapeg}
g({\bf x}) = 0
\end{equation}
\end{subequations}
In passing we note that in the case of uniaxial hard ellipsoids, we have $f({\bf x},t) = {}^{t}({\bf x}- {\bf r}_A(t)) X_A(t) ({\bf x}- {\bf r}_A(t)) - 1$ and a similar expression holds for $g({\bf x})$, where ${\bf r}_A$ is the position 
of the center-of-mass of $A$ and $X_A(t)$ is a matrix, that depends on the orientation of the  $A$ 
(see \cite{DonevTorqStill2}).  
It is assumed that, if a point ${\bf x}$ is inside the rigid body $A$ ($B$), then $f({\bf x}) < 0$  ($g({\bf x}) < 0$),  while if it is outside $f({\bf x}) > 0$ ($g({\bf x}) > 0$). 
Then the distance $d$ between these two objects can be defined as follows:
\begin{equation}\label{Eq:distdef}
d = \min_{ \genfrac{}{}{0pt}{}{f({\bf x}_A)=0}{g({\bf x}_B)=0} } \| {\bf x}_A - {\bf x}_B \|  
\end{equation}
Equivalently, the distance between these two objects can be defined as the solution of the following set of equations:
\begin{subequations} \label{Eq:dist8}
\begin{equation} \label{Eq:dist8a}
f_{{\bf x}_A} = - \alpha^2 g_{{\bf x}_B}
\end{equation}
\begin{equation} \label{Eq:dist8b}
f({\bf x}_A) = 0
\end{equation}
\begin{equation}\label{Eq:dist8c}
g({\bf x}_B) = 0
\end{equation}
\begin{equation} \label{Eq:dist8d}
 {\bf x}_A + \beta f_{{\bf x}_A}={\bf x}_B
\end{equation}
\end{subequations}
where ${\bf x}_A = (x_A,y_A,z_A)$, ${\bf x}_B = (x_B,y_B,z_B)$,
$f_{{\bf x}_A} = \PD{f}{{\bf x}_A}$ and $g_{{\bf x}_B} = \PD{f}{{\bf x}_B}$.
Eqs. (\ref{Eq:dist8b}) and (\ref{Eq:dist8c}) ensure that ${\bf x}_A$ and ${\bf x}_B$ are points on $A$ and $B$, eq.(\ref{Eq:dist8a}) provides that the normals to the surfaces are anti-parallel, and eq.(\ref{Eq:dist8d}) ensures that the displacement of ${\bf x}_A$ from ${\bf x}_B$ is collinear to the normals of the surfaces. 
Equations (\ref{Eq:dist8}) define extremal points of $d$; therefore, for two general HB these equations can have multiple solutions, while only the smallest one is the actual distance. To solve such equations iteratively is therefore necessary to start from a good initial guess of $\left( {\bf x}_A,{\bf x}_B,\alpha,\beta \right )$ to avoid finding spurious solutions.


In addition note that if two HB overlap slightly (i.e. the overlap volume is small) there is a solution with $\beta < 0$ that is a measure of the inter-penetration of the two rigid bodies; we will refer to such a solution as the ``negative distance'' solution. 

\subsubsection{Newton-Raphson method for the distance}
The set of equations (\ref{Eq:dist8}) can be conveniently solved by a Newton-Raphson (NR)
method, as long as first and second derivatives of $f({\bf x}_A)$ and $g({\bf x}_B)$
are well defined. This method, provided that a good initial guess has been supplied, very quickly reaches the solution because of its quadratic convergence \cite{NumRecipes}. If we define:
\begin{equation}
{\bf F}({\bf y})=
\begin{pmatrix}
f_{{\bf x}_A} + \alpha^2 g_{{\bf x}_B}\cr
f({\bf x}_A)\cr
g({\bf x}_B)\cr
{\bf x}_A + \beta f_{{\bf x}_A}-{\bf x}_B
\end{pmatrix}
\end{equation}
Eqs. (\ref{Eq:dist8}) become:
\begin{equation}\
{\bf F}({\bf y})= 0
\end{equation}
where ${\bf y}=({\bf x}_A, {\bf x}_B, \alpha, \beta)$.

Given an initial point ${\bf y}_0$ we build a sequence of points converging to the solution as follows:
\begin{equation}
{\bf y}_{i+1} = {\bf y}_i + {\bf J}^{-1} {\bf F}({\bf y}_i)
\label{Eq:NRiter}
\end{equation}
where ${\bf J}$ is the Jacobian of ${\bf F}$,i.e. ${\bf J} = \PD{\bf F}{\bf y}$.

The matrix inversion, required to evaluate ${\bf J}^{-1}$, can be done making use of a standard $LU$  decomposition \cite{NumRecipes}.  This decomposition is of order $N^3/3$, where is $N$ the number of equations ($8$ in the present case).
Finally we note that the set of $8$ equations (\ref{Eq:dist8}) can be also reduced to $5$ equations, 
eliminating ${\bf x}_A$ or ${\bf x}_B$, using Eq. (\ref{Eq:dist8d}) .


\subsection{Prediction of the time-of-collision}
The collision (or contact) time of two rigid bodies $A$ and $B$ is, the smallest time $t_c$,
such that $d(t_c)=0$, where $d(t)$ is the distance as a function of time between $A$ and $B$.
For finding $t_c$ we perform the following steps:
\begin{enumerate}
\item Bracket the contact time using {\it centroids} (this technique will not be discussed here, see \cite{DonevTorqStill1,DonevTorqStill2} for the details)\footnote{A centroid of a given ellipsoid $A$ with center $\BS r_A$ is the smallest sphere centered at $\BS r_A$ that  that encloses $A$}
\item Overestimating the rate of variation of the distance with respect to time ($\dot d(t)$), refine the bracketing 
of the collision time obtained in $1$.
\item Find the collision time to the best accuracy using a Newton-Raphson on a suitable set of equations for the contact
point and the contact time.\footnote{Alternatively you can use any one-dimensional root-finder for the equation $d(t)=0$.}  
\end{enumerate}


\subsubsection{Bracketing of the contact time overestimating $\dot d(t)$.}
It can be proved that an overestimate of the rate of variation of the distance is the following:
\begin{equation}
\dot d(t) 
\le \| {\bf v}_A - { \bf v}_B \| +  \| { \bf w}_A \| L_A +
\| { \bf w}_B \| L_B
\label{Eq:distOver}
\end{equation}
where the dot indicates the derivation with respect to time, ${ \bf r}_A$ and ${\bf r}_B$ are the centers of mass of the two rigid bodies, ${\bf v}_A$ and ${\bf v}_B$ are the velocities of the centers of mass,  
and the lengths $L_A$, $L_B$ are such that
\begin{subequations}
\label{Eq:lalb}
\begin{equation}
L_A \ge \max_{{f({\bf r}')\le 0}} \{\|{\bf r}'-{\bf r}_A\|\} 
\end{equation}
\begin{equation}
L_B \ge \max_{{g({\bf r}')\le 0}} \{ \|{\bf r}'-{\bf r}_B\|\} 
\end{equation}
\end{subequations}

Using this overestimate of $\dot d(t)$, that will be called $\dot d_{max}$ from now on, an efficient strategy, to refine the bracketing of the contact time, is the following:

\begin{enumerate}
\item If $t_1$ and $t_2$ bracket the solution, set $t=t_1$.
\item Evaluate the distance $d(t)$ at time $t$.
\item Choose a time increment $\Delta t$ as follows:
\begin{equation}	
\Delta t =  
\begin{cases}
\frac{d(t)}{\dot d_{max}}, & \hbox{if}\> d(t) > \epsilon_d \hbox{;} \cr
\frac{\epsilon_d}{\dot d_{max}}, & \hbox{otherwise.} \cr
\end{cases}
\end{equation}
where $\epsilon_d \ll \min\{L_A,L_B\}$
\item Evaluate the distance at time $t+\Delta t$.
\item If $d(t+\Delta t) < 0$ and $d(t) > 0$, then $t_1=t$ and $t_2=t+\Delta t$, find the collision time/point via NR (see Sec. \ref{Subsec:lastrefine}) and terminate (collision will occur after $t_1$).
\item if both $0 < d(t+\Delta t) < \epsilon_d$ and $0 < d(t) < \epsilon_d$, there could be a ``grazing`` collision between $t$ and $t+\Delta t$ \cite{DonevTorqStill2} (distance is first decreasing and then increasing). To look for a possible collision, evaluate the distance $d(t+\Delta t /2)$ and perform a quadratic interpolation of the $3$ points ( $(t,d(t))$, $(t+\Delta t/2, d(t+\Delta t/2)$, $(t+\Delta t, d(t+\Delta t)$ ). If the resulting parabola has zeros, set $t_1$ to the smallest zero (first collision will occur near the smallest zero), and find the collision time/point via NR and terminate (see Sec. \ref{Subsec:lastrefine}). 
\item Increment time by $t\rightarrow t+\Delta t$.
\item if $t > t_2$ terminate (no collision has been found).
\item Go to step 2.
\end{enumerate}
Finally we note that if the quadratic interpolation fails, the collision will be missed, anyway if $\epsilon_d$ is enough small all ``grazing`` collisions will be properly handled, i.e. all collisions will be correctly predicted. 

\subsubsection{Set of equations to find the contact time and the contact point}
\label{Subsec:lastrefine}
The contact time, to the best possible accuracy, can be found  
solving the following equations:
\begin{subequations} \label{Eq:contime}
\begin{equation} \label{Eq:contimea}
f_{\BS x}(\BS x, t) = -\alpha^2 g_{\BS x}(\BS x,t)
\end{equation}
\begin{equation} \label{Eq:contimeb}
f(\BS x, t) = 0
\end{equation}
\begin{equation} \label{Eq:contimec}
g(\BS x, t) = 0
\end{equation}
\end{subequations}
where now $f$ and $g$ depend also on time because the two objects move, and the independent variables are the contact time
and the contact point.
Again, a good way to solve such a system is using the NR method, very similarly to what we did for evaluating the geometric distance.
NR for Eqs. (\ref{Eq:contime}) is again very unstable unless a very good initial guess is provided, but
the bracketing evaluated using $\dot d_{max}$ is sufficiently accurate, provided that $\epsilon_d$ is enough small (typically
$\epsilon_d \lessapprox 1E-4$ is a good choice to give a good initial guess for this NR and to avoid ``grazing`` collisions).  

Finally we want to stress that this algorithm will be exploited fully, when hard bodies of arbitrary shape will be simulated, and that with minor changes it can also be used to simulate hard bodies decorated with attractive spots, as it has been done in the past for the specific case of hard-spheres \cite{DeMicheleH2O,DeMicheleSiO2}.  
Work along these directions is on the way, and in particular a system of super-ellipsoids
has been already successfully simulated using the present algorithm. 
\section{Methods}
\label{Sec:methods}
We perform an extended study of the dynamics of monodisperse HE
in a wide window of $\phi$ and $X_0$ values, extending the range of $X_0$
previously studied \cite{AllenFrenkelDyn}. We specifically focus on establishing the
trends leading to dynamic slowing down in both translations and rotations,
by  evaluating  the loci of constant translational and rotational diffusion. These lines, in the limit of vanishing diffusivities, approach the glass-transition lines.
We also study translational and rotational correlation functions, to search for
the onset of slowing down and stretching in the decay of the correlation. 
We simulate a system of $N=512$ ellipsoids at various volumes $V=L^3$ in a cubic box of edge $L$ with periodic boundary conditions.  
We chose the geometric mean of the axis $l=\sqrt[3]{ab^{2}}$ as unit
of distance, the mass $m$ of the particle as unit of mass ($m=1$) and
 $k_BT=1$ (where  $k_{B}$ is the Boltzmann constant and $T$ is the temperature) and hence  the corresponding unit of time is $\sqrt{ml^{2}/k_{B}T}$.
The inertia tensor is chosen as $I_{x}=I_{y}= 2mr^{2}/5$, where $r=\min\{a,b\}$. The value of the $I_{z}$ component 
is irrelevant \cite{allenfrenkelBook}, since  the angular velocity along the symmetry (z-) axis of the HE is conserved. 
We simulate a grid of more than 500 state points at different $X_0$ and $\phi$.
To create the starting configuration at a desired $\phi$, we generate a random distribution of ellipsoids at very low $\phi$ and then we progressively  decrease $L$ up to the desired $\phi$. 
We then equilibrate the configuration by propagating the trajectory for times such that 
both angular and translational correlation functions have decayed to zero. 
Finally, we perform a production run at least $30$ times longer than the time needed to equilibrate. For the points close to the I-N transition we check the nematic order by evaluating the largest eigenvalue $S$ of the order tensor ${\bf Q}$~\cite{SpheroCylRec}, whose components are: 
\begin{equation}
Q_{\alpha\beta} = \frac{3}{2}\frac{1}{N}\sum_i \langle({\bf u}_i)_{\alpha} ({\bf u}_i)_{\beta}\rangle - \frac{1}{3} \delta_{\alpha,\beta}
\end{equation}
where $\alpha\beta\in\{x,y,z\}$, and the unit vector $({\bf u}_i(t))_\alpha$ is the component $\alpha$ 
of the orientation (i.e. the symmetry axis) of ellipsoid $i$ at time $t$.
The largest eigenvalue $S$ is non-zero if the system is nematic and $0$
if it is isotropic.  In the following, we choose  the value $S = 0.3$ as criteria to separate isotropic from
nematic states. 

\section{Results and discussion}
\label{Sec:results}
\subsection{Isodiffusivity lines}
\begin{figure}
\vspace{1cm}
\includegraphics[width=.45\textwidth]{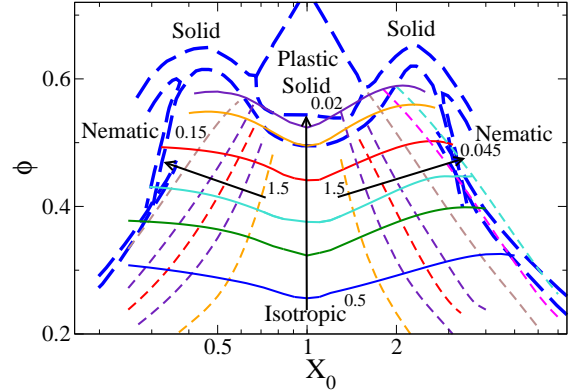}
\caption{[Color online] Isodiffusivity lines. Solid lines are isodiffusivity lines from translational diffusion coefficients
$D_{trans}$ and dashed lines are isodiffusivities lines from rotational diffusion coefficients $D_{rot}$.
Arrows indicate decreasing diffusivities. Left and right arrows refer to rotational diffusion 
coefficients. Diffusivities along left arrow are: $1.5$, $0.75$, $0.45$, $0.3$, $0.15$.
Diffusivities along right arrow are: $1.5$, $0.75$, $0.45$, $0.3$, $0.15$, $0.075$, $0.045$. 
Central arrow refers to 
translational diffusion coefficients, whose values are: $0.5$, $0.3$, $0.2$, $0.1$, $0.04$, $0.02$.
Thick Long-dashed curves are coexistence curves of all first order phase transitions in the phase diagram of HE
evaluated by Frenkel and Mulder (FM) \cite{FrenkelPhaseDiagMolPhys}
Solid lines are 
coexistence curves for the I-N transition of oblate and prolate ellipsoids, 
obtained analytically by Tijpto-Margo and Evans \cite{MargoEvans} (TME).}
\label{Fig:isod}
\end{figure}

From the grid of simulated state points we build a corresponding grid of
translational  ($D_{trans}$) and diffusional  ($D_{rot}$) coefficients, defined as:
\begin{equation}
D_{trans} = \lim_{t\rightarrow+\infty}  \frac{1}{N} \sum_i \frac{\langle \|{\bf x}_i(t) - {\bf x}_i(0)\|^2  \rangle}{6 t}  
\end{equation}
\begin{equation}
D_{rot} =  \lim_{t\rightarrow+\infty}  \frac{1}{N} \sum_i \frac{\langle \| \Delta\Phi_i \|^2  \rangle}{4 t}  
\end{equation}
where $\Delta\Phi_i = \int_0^t {\bf \omega}_i dt$,
${\bf x}_i$ is the position of the center of mass and $ {\bf \omega}_i$  is the angular velocity of ellipsoid $i$.
By proper interpolation, we evaluate the 
isodiffusivity lines,  shown in Fig. \ref{Fig:isod}. 
Results show a striking decoupling of the translational and rotational
dynamics. While the translational isodiffusivity lines 
mimic the swallow-like shape  of the coexistence between the isotropic liquid and the crystalline phases (as well as the MMCT prediction for the glass transition \cite{LetzSchilLatz}), rotational isodiffusivity lines reproduce qualitatively the shape of the I-N coexistence. 
As a consequence of the the swallow-like shape, at large fixed $\phi$, $D_{trans}$   increases by increasing the particle's  anisotropy, reaching its maximum at  $X_0\approx 0.5$ and $X_0\approx 2$. 
Further increase of the anisotropy results in a decrease of $D_{trans}$.  For all $X_0$, an increase of $\phi$ at constant $X_0$ leads to  a significant suppression of $D_{trans}$, demonstrating that $D_{trans}$ is controlled
by packing.
The iso-rotational lines are instead mostly controlled  by $X_0$, showing a progressive slowing down of the rotational dynamics independently from the translational behavior. This suggests that on moving along a path of
constant $D_{trans}$, it is possible to progressively decrease the rotational dynamics, up to the point where rotational diffusion arrest and all rotational motions become hindered.
Unfortunately, in the case of monodisperse HE, a
nematic transition intervenes well before this point is reached. It is thus stimulating to think about the possibility of designing a system of hard particles in which the nematic transition is inhibited by a proper
choice of the disorder in the particle's shape/elongations. 
We note that the slowing down of the rotational dynamics is consistent with MMCT predictions of a nematic glass for large $X_0$ HE \cite{LetzSchilLatz}, in which  orientational degrees of freedom start to freeze approaching the isotropic-nematic transition line, while translational degrees of freedom mostly remain ergodic.
\subsection{Orientational correlation function}
To support the possibility that the slowing down of the dynamics 
on approaching the nematic phase originates from a close-by glass transition,
we evaluate  the self part of the intermediate scattering function 
$F_{self}(q,t) = \frac{1}{N} \langle \sum_j e^{i{\bf q}\cdot({\bf x}_j(t) - {\bf x}_j(0))} \rangle$
and  the second order orientational correlation function
 $C_2(t)$ defined as \cite{AllenFrenkelDyn} 
$C_2(t) = \langle P_2(\cos\theta(t))\rangle $,
where $P_2(x) = (3 x^2 - 1) / 2$ and $\theta(t)$ is the angle between the symmetry axis at time $t$ and at time $0$.
The $C_2(t)$ rotational isochrones  
are found to be very similar to rotational isodiffusivity lines.  

These two correlation functions never show a clear two-step relaxation decay in the entire 
studied  region, even  where the isotropic phase is metastable, since the system can not be
significantly over-compressed. 
As for the well known hard-sphere case, the amount of over-compressing 
achievable in a monodisperse system is rather limited.  This notwithstanding, a comparison of
the rotational and translational correlation functions reveals that the onset of dynamic slowing down and glassy 
dynamics can be detected by the appearance of stretching. 

We note that $F_{self}$ shows an exponential behaviour close to the I-N transition ($X_0=3.2$,$\ 0.3448$) 
on the prolate and oblate side,  in agreement
with the fact that translational isodiffusivities lines do not exhibit any peculiar behaviour close to the I-N line 
\cite{DeMichelePRL2007}. Only when $X_0\approx 1$, 
$F_{self}$ develops a small stretching, consistent with the minimum of the swallow-like curve observed in the 
fluid-crystal line \cite{HardSpheresExp,HardSpheresSim},  in the jamming locus as well as  in the predicted behavior of the glass line for HE \cite{LetzSchilLatz} and for small elongation dumbbells \cite{DumbellChongGoetze,DumbellChongFra}.  Opposite behavior is seen for the case of the 
orientational correlators. $C_2$ shows stretching at large anisotropy, i.e. at small and large $X_0$ values,
but decays within the microscopic time for almost spherical particles. 
In this quasi-spherical limit, the decay is well represented by the decay of  a free rotator \cite{FreeRotator,DeMichelePRL2007}. 
Previous studies of the rotational dynamics of 
HE \cite{AllenFrenkelDyn} did not report  stretching in $C_2$, probably due to the smaller values of $X_0$ previously investigated and to the present increased statistic  which allows us to follow the full decay of the correlation functions.

\begin{figure}
\vspace{1cm}
\includegraphics[width=.45\textwidth]{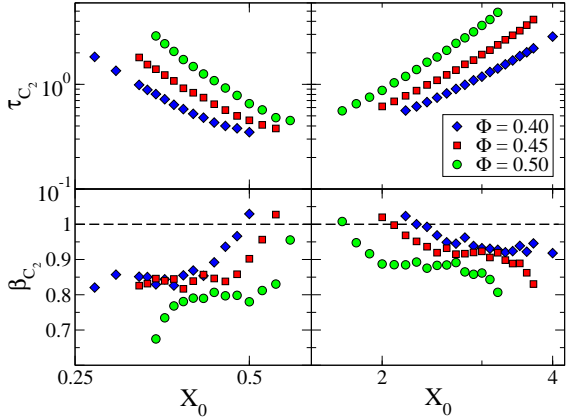}
\caption{[Color online] $\beta_{C_2}$ and $\tau_{C_2}$ are obtained from fits of $C_2$ to a stretched exponential for $\phi=0.40,0.45$ and $0.50$. Top: $\tau_{C_2}$ as a function 
of $X_0$. Bottom: $\beta_{C_2}$ as a function of $X_0$. The time window used 
for the fits is chosen in such a way to exclude the microscopic short times ballistic
relaxation. For $0.588 < X_0 < 1.7$ the orientational relaxation is 
exponential.}
\label{Fig:betavsX0}
\end{figure}

In summary $C_2$ becomes stretched approaching the I-N transition while $F_{self}$  remains exponential on approaching the  transition.
To quantify the amount of stretching in $C_2$, we fit it to the function $A \exp[-(t/\tau_{C_2})^{\beta_{C_2}}]$ (stretched exponential) for several state points and 
we show  in Fig. \ref{Fig:betavsX0}
the $X_0$ dependence of $\tau_{C_2}$ and $\beta_{C_2}$ for three different values of $\phi$.
In all cases, slowing down of the characteristic time and stretching increases progressively on approaching the I-N transition. 
\section{Conclusions}
\label{Sec:conclusions}
In summary we have applied a novel algorithm for simulating hard bodies to investigate the dynamics properties of a system of
monodisperse HE and we have shown that clear precursors of dynamic slowing down and stretching  can be observed in  the region of the  phase diagram where a (meta)stable isotropic phase can be studied. Despite the monodisperse character of the present system prevents the possibility of observing a clear glassy dynamics, 
our data  suggest that a slowing down in the orientational degrees of freedom ---  driven by the elongation of the particles --- is in action. The main effect of this shape-dependent slowing down is 
a decoupling of the translational and rotational dynamics which generates an almost perpendicular crossing of the $D_{trans}$ and $D_{rot}$ isodiffusivity lines. 
This behavior is in accordance with  MMCT predictions, suggesting two glass transition mechanisms, related respectively to cage effect (active for $0.5 \lessapprox X_0 \lessapprox 2$) and to pre-nematic order ($X_0 \lessapprox 0.5$, $X_0 \gtrapprox 2$) \cite{LetzSchilLatz}. 
\begin{theacknowledgments}
We acknowledge support from MIUR-PRIN. 
\end{theacknowledgments}



\bibliographystyle{aipproc}   

\bibliography{DeMichele}

\IfFileExists{\jobname.bbl}{}
 {\typeout{}
  \typeout{******************************************}
  \typeout{** Please run "bibtex \jobname" to optain}
  \typeout{** the bibliography and then re-run LaTeX}
  \typeout{** twice to fix the references!}
  \typeout{******************************************}
  \typeout{}
 }

\end{document}